\documentclass[usenatbib,usegraphicx]{mn2e}

\title{Cosmic Voids in Sloan Digital Sky Survey Data Release 7}

\author[D. Pan et al.]{Danny C. Pan$^{1}$, Michael S. Vogeley$^{1}$, Fiona Hoyle$^{2}$,Yun-Young Choi$^{3}$, Changbom Park$^{4}$ \\
$^{1}$Department of Physics, Drexel University, Philadelphia, PA 19104 \\
$^{2}$Department of Physics and Astronomy, Widener University, Chester, PA 19013 \\
$^{3}$Astrophysical Research Center for the Structure and Evolution of the Cosmos, Sejong University, Seoul 143-747, Korea \\
$^{4}$Korea Institute for Advanced Study, Seoul 130-722, Korea}

\begin{document}

\maketitle

\begin{abstract} 
We study the distribution of cosmic voids and void galaxies using Sloan Digital Sky Survey Data Release 7 (SDSS DR7).  Using the VoidFinder algorithm as described by \citet{Hoyle:2002}, we identify 1054 statistically significant voids in the northern galactic hemisphere with radii $> 10 h^{-1}$ Mpc.  The filling factor of voids in the sample volume is 62\%.  The largest void is just over $30 h^{-1}$ Mpc in effective radius.  The median effective radius is $17 h^{-1}$ Mpc.  The voids are found to be significantly underdense, with density contrast $\delta < -0.85$ at the edges of the voids.  The radial density profiles of these voids are similar to predictions of dynamically distinct underdensities in gravitational theory.  We find 8,046 galaxies brighter than $M_r = -20.09$ within the voids, accounting for 7\% of the galaxies.  We compare the results of VoidFinder on SDSS DR7 to mock catalogs generated from a SPH halo model simulation as well as other $\Lambda$-CDM simulations and find similar void fractions and void sizes in the data and simulations.  This catalog is made publicly available at http://www.physics.drexel.edu/$\sim$pan/VoidCatalog for download.
\end{abstract}


\section{Introduction}

Redshift surveys of galaxies reveal a rich variety of large-scale
structures in the Universe: clusters that span a few megaparsecs in
radius, connected by filaments stretching up to many tens of
megaparsecs, which in turn envelop vast underdense voids with radii of
tens of megaparsecs. These large scale structures are described by
\citet{Bond:1996} as a Cosmic Web of material that reflects the
initial density fluctuations of the early Universe. While historically
most attention has been paid by astronomers to the dense clusters and
filaments, it is the voids that fill most of the volume in the
Universe.  These underdense regions strongly influence the growth of
large scale structure. The statistics and dynamics of cosmic voids and
the properties of the few objects found within them provide critical tests
of models of structure formation.

Observations of voids in the galaxy distribution have progressed as
the depth, areal coverage, and sampling density of galaxy redshift
surveys has improved.  \citet{Rood:1988} reviews the paradigm shift
that occured beginning in the mid-1970s as the focus shifted from the
study of galaxy surface distributions to three-dimensional spatial
distributions provided by redshift surveys, and the impact of this
revolution on studies of voids.  \citet{Joeveer:1978} identified
superclusters and voids in the distribution of galaxies and Abell
clusters.  Pencil beam surveys of the Coma/Abell 1367 supercluster
\citep{Gregory:1978} indicated large voids.  \cite{Kirshner:1981}
discovered a void in the Bootes region of the sky that is 50 $h^{-1}$
Mpc in diameter, several times larger than any previously observed.
The Center for Astrophysics Redshift Survey \citep{Huchra:1983} and in
particular its extension to $m_B=15.5$ \citep{deLapparent:1986,
Geller:1989} revealed that the large-scale structure of galaxies is
dominated by large voids and the sharp filaments and walls that
surround them.  The Southern Sky Redshift Survey
\citep{daCosta:1988,Maurogordato:1992} found similar results.  The
\citet{Giovanelli:1985} survey detailed the supercluster and void
structure of the Perseus-Pisces region.  The deeper Las Campanas
Redshift Survey \citep{Kirshner:1991,Shectman:1996} confirmed the
ubiquity of voids in the large-scale distribution of galaxies.
Comparison of optically-selected galaxy surveys with redshift surveys
of infrared selected galaxies
\citep{Strauss:1992,Fisher:1995,Saunders:2000,Jones:2004} indicated
that the same voids are found regardless of galaxy selection.  The
completed Two Degree Field Galaxy Redshift Survey (2dFGRS;
\citet{Colless:2001}) and Sloan Digital Sky Survey (SDSS;
\citet{York:2000, Abazajian:2009}) now allow the most complete view
to date of the detailed structure of voids.


A variety of methods have been used to compile catalogs of voids in both observations of galaxies or clusters and in simulations (using dark matter particles or mock galaxy catalogs). Detailed discussion of many of these methods is given by 
\citet{Colberg:2008},  who compare void finding techniques. For the purpose of finding voids in redshift survey observations,  methods that are applicable to the distribution of galaxies include 
\citet{Kauffmann:1991,
El-Ad:1997b,
Aikio:1998,
Hoyle:2002,
Neyrinck:2008,
Aragon-Calvo:2010}.
Examples of applications of such methods to galaxy redshift surveys include analyses of
the Southern Sky Redshift Survey
\citep{Pellegrini:1989}, the first slice of the Center for Astrophysics Redshift Survey
\citep{Slezak:1993},
as well as the full extension of the CfA Redshift Survey
\citep{Hoyle:2002},
 the IRAS 1.2Jy and Optical Redshift Surveys
\citep{El-Ad:1997a, El-Ad:1997b, El-Ad:2000}, the Las Campanas Redshift Survey
\citep{Muller:2000}, the IRAS PSCz Survey 
\citep{Hoyle:2002, Plionis:2002}, 
the 2dFGRS
\citep{Hoyle:2004, Ceccarelli:2006, Tikhonov:2006a},
and preliminary data from the SDSS
\citep{Tikhonov:2007, Foster:2009}.





The importance of cosmic voids as dynamically-distinct elements of
large-scale structure is clearly established by theory
\citep{Hoffman:1982,Hausman:1983,Fillmore:1984,
Icke:1984,Bertschinger:1985,Blumenthal:1992,Sheth:2004,Patiri:2006c,Furlanetto:2006}.  Linear
theory predicts that the interior of the voids should reach a flat
plateau and the boundaries of the voids should be quite sharp.
Simulations of structure formation (e.g.,
\citet{Regos:1991,Dubinski:1993,vandeWeygaert:1993,Colberg:2005})
demonstrate that large voids are caused by super-Hubble outflows that
are nearly spherically symmetric out to near the edges of the
voids. Tidal effects of clusters only become important for objects
near the walls around voids.  Simulations of the Cold Dark Matter
model for structure formation indicate that the interiors of voids
should include dark matter filaments and many low mass halos
\citep{Mathis:2002,Benson:2003a,Gottlober:2003}. Identifying these
structures is an important test of this model.

The properties of large voids in the distribution of galaxies may provide strong tests of cosmology. 
\citet{Ryden:1995, Ryden:1996} discuss the use of void shapes in redshift space as a cosmological test. More recent work examines voids as a probe of dark energy 
\citep{Park:2007,
Lee:2009,
Biswas:2010,
Lavaux:2010}
Comparison of voids at low and high redshift may provide a strong test of the $\Lambda$CDM model
\citep{Viel:2008}.
The abundance of cosmic voids is a critical probe for non-gaussianity in the initial conditions for structure formation
\citep{Kamionkowsi:2009,
Chongchitnan:2010}.
Beyond tests of the $\Lambda$CDM model, the properties of voids galaxies may even constrain alternative theories gravity
\citep{Hui:2009}.












Mapping the voids is important both for studying large-scale cosmic
structure and because they are a unique astrophysical laboratory for
studying galaxy formation. Gravitational clustering within a void
proceeds as if in a very low density universe, in which structure
formation occurs early and there is little interaction between
galaxies, both because of the lower density and the faster local
Hubble expansion. Goldberg \& Vogeley (2004) show that the interior of
a spherical void with 10\% of the mean density in a flat
$\Omega_{matter}=0.3$ $h=0.7$ universe evolves dynamically like an
$\Omega_{matter}=0.02$, $\Omega_{\Lambda}=0.48$, $h=0.84$ universe.

\citet{Peebles:2001} describes the ``void phenomenon": galaxies of all
types appear to respect the same voids, in contrast to the prediction
of CDM that low density regions should contain many low mass objects.
\citet{Tikhonov:2009} find, using comparison of voids and void galaxies in the local volume with high-resolution simulations, that the emptiness of voids is a problem for $\Lambda$CDM.

Voids are expected to harbor many low mass halos that are the ideal
breeding grounds for faint galaxies; if the low mass halos predicted
by CDM harbor luminous galaxies, then they should be optically
visible.  Optical observations have not revealed a large population of
fainter galaxies in voids \citep{Thuan:1987,Linder:1996,Kuhn:1997,Popescu:1997},
although the luminosity function in voids is shifted by about one full
magnitude \citep{Hoyle:2005}.  
\citet{Tinker:2009} contends the $\Lambda$CDM void phenomenon is due
to a lack of understanding of assembly bias as galaxies form,
but their model predicts a 5-magnitude shift in maximum galaxy luminosity.
If void halos contain gas, but too few stars
to be visible, then then their gas might be detected.  To date, blind
HI surveys have not detected such a population of HI rich but
optically dark galaxies \citep{Haynes:2008}. 
Nearby Lyman-$\alpha$ clouds detected along lines of sight toward bright quasars show a strong preference for inhabiting the voids, but most of these clouds seem to be associated with galaxy structures
(Pan et al., in preparation).

In contrast to a picture in which star formation in void halos is suppressed, our analyses of void galaxies in the SDSS DR2 and DR4 samples indicate that void galaxies are bluer and have higher specific star formation rates than galaxies in denser environments 
\citep{Rojas:2004, Rojas:2005, CPark:2007}.  
For the small number of dwarf galaxies in the earlier samples, we note even stronger trends with environment; at fixed morphology and luminosity, the faintest void galaxies are bluer and have higher star formation rates.  Focusing on the blue population in voids, we find in these preliminary SDSS analyses, and 
\citet{vonBenda-Beckmann:2008} find in 2dFGRS, that this blue population is not only more numerous, but also bluer and with higher star formation than in denser regions.


In \citet{Hoyle:2005} we find a much fainter exponential cutoff in the luminosity function in voids ($\Delta M_r^* = 1.1$ mag) but no evidence for a change in the faint end slope between voids and ``walls.''  However, the uncertainties at faint magnitudes are quite large.  We could not find a sub-population of ``wall'' galaxies selected by color, surface brightness profile, or $H\alpha$ equivalent width that matched both the faint end slope $\alpha$ and characteristic magnititude $M_r^*$ of void galaxies.  In \citet{CPark:2007} we again find that $M^*$ monotonically shifts fainter at lower density.  and that the faint end (measured only down to $M_r=-18.5$) slope varies significantly with density. These results are consistent with earlier analyses \cite{Grogin:1999, Grogin:2000}.  These trends also persist into the ``wall'' regions closest to large voids \citep{Ceccarelli:2008}. When we estimate the mass function of void galaxies in SDSS and compare to the luminosity function \citep{Goldberg:2005}
we find a good match with the predictions of the Press-Schechter model, thus the void galaxies appear to be nearly unbiased with respect to the mass. 

While we see some clear trends, controversy persists in the literature as to whether or not galaxies in voids differ in their internal properties from similar objects in denser regions. For example, 
\citet{Rojas:2004, Rojas:2005, Blanton:2005b, Patiri:2006b}, and \citet{vonBenda-Beckmann:2008} reach varying conclusions that clearly depend on how environment is defined and which observed properties are compared.  There is a marked difference between properties of the least dense 30\% of galaxies (in regions with density contrast $\delta<-0.5$) and objects in deep voids which form the lowest density 10\% of galaxies (in regions of density contrast $\delta<-0.8$, which is the theoretical prediction for the interiors of voids that are now going non-linear). 
All of these results, and the possible controversy among them, highlight the importance of building the largest possible, publicly released catalog of voids and void galaxies.

Lastly, for the purpose of examining the influence of environment on galaxy
formation and evolution, it is important to make a distinction between
void galaxies and isolated galaxies.  Void galaxies are galaxies that
reside within large scale void structures in the Universe.  While this
has an overall effect on the local environments of these void
galaxies, it does not preclude galaxies from residing within small
scale dense environments, or cloud-in-void as described in
\citet{Sheth:2004}.  Isolated galaxies are generally found by nearest
neighbor distance measures typically on the scale of small (Mpc)
nearby environments \citep{Karachentsev:2010,Karachentseva:1973}; they
do not necessarily reside in large scale voids.


The purpose of this paper is to obtain the largest void catalog available to date, for the purpose of allowing precision cosmological tests with voids and more accurate tests of galaxy formation theories.  We utilize a galaxy based void finding algorithm, ``VoidFinder'' \citep{Hoyle:2002}, to identify voids in the final galaxy catalog from SDSS (DR7).  This void finding technique is shown to accurately identify large-scale cosmic voids with properties similar to those predicted by gravitational instability theory.  Section 2 describes the VoidFinder algorithm.  Section 3 describes the SDSS data used for this research.  Section 4 presents results on the various properties of the voids found.  Sections 5 and 6 describe several methods used to test the robustness of the method.


\section{VoidFinder}
VoidFinder is a galaxy-based void finding algorithm that uses redshift data to find statistically significant cosmic voids.  VoidFinder is based on the \citet{El-Ad:1997b} method as implemented by \citet{Hoyle:2002} and uses a nearest neighbor algorithm on a volume limited galaxy catalog for void finding. This approach is highly effective in identifying large voids of density contrast $\delta\le -0.9$ and radius $R> 10h^{-1}$Mpc. 
The method is robust when applied to different surveys that cover the same volume of space (we have applied VoidFinder to IRAS PSCz, CfA2+SSRS2, 2dF, SDSS, 6dF and compared overlaps; see \citet{Hoyle:2002,Hoyle:2004}).
Our tests on cosmological simulations demonstrated that this method works in identifying voids in the distributions of both simulated galaxies and dark matter \citep{Benson:2003a}. 

VoidFinder is applied to volume limited galaxy samples.  The galaxies are initially classified as wall or field galaxies.  A field galaxy is a galaxy that may live in a void region whereas wall galaxies lie in the cosmic filaments and clusters.  The distance parameter $d$ for determining whether a galaxy is a wall or field galaxy is based on the third nearest neighbor distance ($d_3$) and the standard deviation of the distance ($\sigma_{d_3}$):
\begin{displaymath}
d = d_3 + 1.5\sigma_{d_3} 
\end{displaymath}
In our galaxy sample, this selection parameter is $d > 6.3 h^{-1}$ Mpc for field galaxies.  With this value of $d$ and choice of $M_{lim} = -20.09$, all field galaxies reside in underdense regions with density contrast $\delta\rho/\rho < -0.47$.  Voids are expected to be significantly underdense, containing approximately 10\% of the cosmic mean density.  Near the edges of the voids, the density is expected to rise very sharply, drastically going from 20\% of the mean density to 100\%.  Using this criterion for the edges of voids, it is expected that the distance criterion for void galaxies will depend on the density at the edge of the void and the spatial correlation of galaxies in voids and the fact that we are sitting on a galaxy.  If we assume that the density at the edge of a void is 20\% of the mean, then the expected density $\rho$ around a galaxy near the void edge can be calculated as
\begin{eqnarray}
\rho(r)/\bar\rho = (0.2)(\xi(r)+1)
\end{eqnarray}
where $\xi(r)$ is the two point autocorrelation function of the galaxy sample.  The average density in a sphere of radius $R$ around a galaxy near the void edge is therefore 
\begin{eqnarray}
\rho(R)/\bar\rho = (0.2)(\bar{\xi(r)}+1)
\end{eqnarray}
where $\bar{\xi}(R)$ is the average value in a sphere of radius $R$.
Over the scales of interest here, the redshift-space correlation function for galaxies in our volume-limited sample can be approximated by a power law,
\begin{eqnarray}
\xi = (s/s_0)^{-\gamma}
\end{eqnarray}
with $s_0$ = 7.62 $\pm$ 0.67 and $\gamma$ = 1.69 $\pm$ 0.1.  Using these values, we can determine the values of $R_{20}$, the radial distance from a void galaxy where we would expect to encounter 20\% of the mean density, and $\delta_{d=6.3}$, the expected underdensity of a void if its third nearest neighbor is found at a distance of 6.3 $h^{-1}$ Mpc away.  We find
\begin{eqnarray}
R_{20} = 4.8^{+0.62}_{-0.74} h^{-1} Mpc \\
\delta_{d=6.3} = -0.88
\end{eqnarray}
We expect that at the edges of the voids $\delta=-0.8$, and in the centers of the voids $\delta=-0.9$.  Thus, our choice for the value of $d$ allows us to pick out void galaxies conservatively, selecting mostly galaxies that only live near the centers of the voids and not allowing void regions to grow into the nonlinear regime.  All galaxies with third nearest neighbor distance $d_3 > 6.3 h^{-1}$ Mpc are considered to be potential void galaxies and are removed from the galaxy sample, leaving us a list of wall galaxies.

We map out the void structure by finding empty spheres in the wall galaxy sample that remains.  Wall galaxies are gridded up in cells of size 5 $h^{-1}$ Mpc, which allows us to find all voids larger than 8.5 $h^{-1}$ Mpc in radius.  All empty cells are considered to be the centers of potential voids.  A maximal sphere is grown from each empty cell, but the center of the maximal sphere is not confined to the initial cell.  Eventually the sphere will be bound by 4 wall galaxies.  There is redundancy in the finding of maximal spheres, but this is useful to define non-spherical voids.

The sample of empty spheres now represents our potential void regions.  We sort the empty spheres by size starting with the largest.  The largest empty sphere is the basis of the first void region.  If there is an overlap of $> 10\%$ between an empty sphere and an already defined void then the empty sphere is considered to be a subregion of the void, otherwise the sphere becomes the basis of a new void.  There is a cutoff of 10 $h^{-1}$ Mpc for the minimum radius of a void region as we seek to find large scale structure voids that are dynamically distinct and not small pockets of empty space created by a sparse sample of galaxies.  Any field galaxies that lie within a void region are now considered void galaxies.  For further details of the VoidFinder algorithm see \citet{Hoyle:2002,Hoyle:2004}.

\section{Data: SDSS DR7}

We use the SDSS Data Release 7 (DR7) \citep{Abazajian:2009} sample of galaxies.  The SDSS is a photometric and spectroscopic survey that covers 8,032 square degrees of the northern sky.  Observations were carried out using the 2.5m telescope at Apache Point Observatory in New Mexico in five photometric bands: u, g, r, i, and z \citep{Fukugita:1996,Gunn:1998}.  Follow up spectroscopy was carried out for galaxies with Petrosian r band magnitude $r < 17.77$ after each photometric image was reduced, calibrated and classified \citep{Lupton:2001,Lupton:1999,Strauss:2002}.

Spectra were taken using circular fiber plugs with an angular size of approximately 55 arc seconds. If two galaxies were closer than this, we could only obtain the spectra of one; the other object is omitted unless there is plate overlap. \citet{Blanton:2003} addresses the issue of fiber collisions by assessing the relation between physical location of the galaxy and photometric and spectroscopic properties and assigns a redshift to the object missed by SDSS.

We use the Korea Institute for Advanced Study Value-Added Galaxy Catalog (KIAS-VAGC) \citep{Choi:2010}.  Its main source is the New York University Value-Added Galaxy Catalog (NYU-VAGC) Large Scale Structure Sample (brvoid0) \citep{Blanton:2005a} which includes 583,946 galaxies with $10 < r \leq 17.6$.  After removing 929 objects that were errors, mostly deblended outlying parts of large galaxies, including 10,497 galaxies excluded by SDSS but that were part of UZC, PSCz, RC3, or 2dF, and also including 114,303 galaxies with $17.6 < m_r < 17.77$ from NYU-VAGC (full0), there is a total of 707,817 galaxies.  This catalog offers an extended magnitude range with high completeness from $10 < r < 17.6$.  There are 120,606 galaxies with $z < 0.107$ and $M_r < -20.09$ in the volume limited sample.


\section{Measurement of Void Properties}

We identify 1,054 voids in SDSS DR7.  The largest voids are 30 $h^{-1}$ Mpc in effective radius, where the volume of the void region is equal to the volume of the sphere with radius $r_{eff}$, and the median effective void radius is 17 $h^{-1}$ Mpc.  The voids cover 62\% of the volume in the sample, and contain 7\% of the volume limited galaxies.  We also identify 79,947 void galaxies with SDSS spectra that lie within the voids in the $r < 17.6$ magnitude limited catalog, which corresponds to 11.3\% of the magnitude limited galaxies.

\begin{figure}
\centering
\includegraphics[scale=0.3,angle=270]{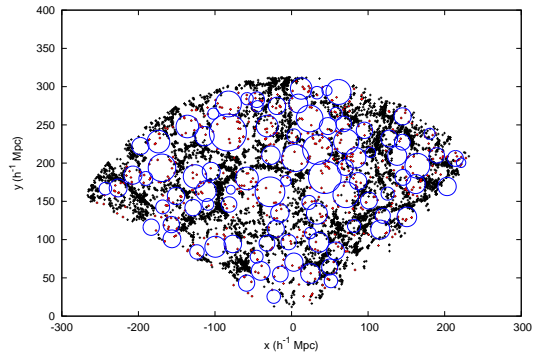}
\caption{10 $h^{-1}$ Mpc thick slab through the middle of the largest void at RA = 226.52960, DEC = 60.41244.  Locations of galaxies ($M_r < -20.09$) are shown with *, and the locations of void galaxies are shown with +.  The circles show the intersection of the maximal sphere of each void with the midplane of the slab.}
\label{redshiftslice}
\end{figure}


\subsection{Void Sizes}
\begin{figure}
\centering
\includegraphics[scale=0.25,angle=270]{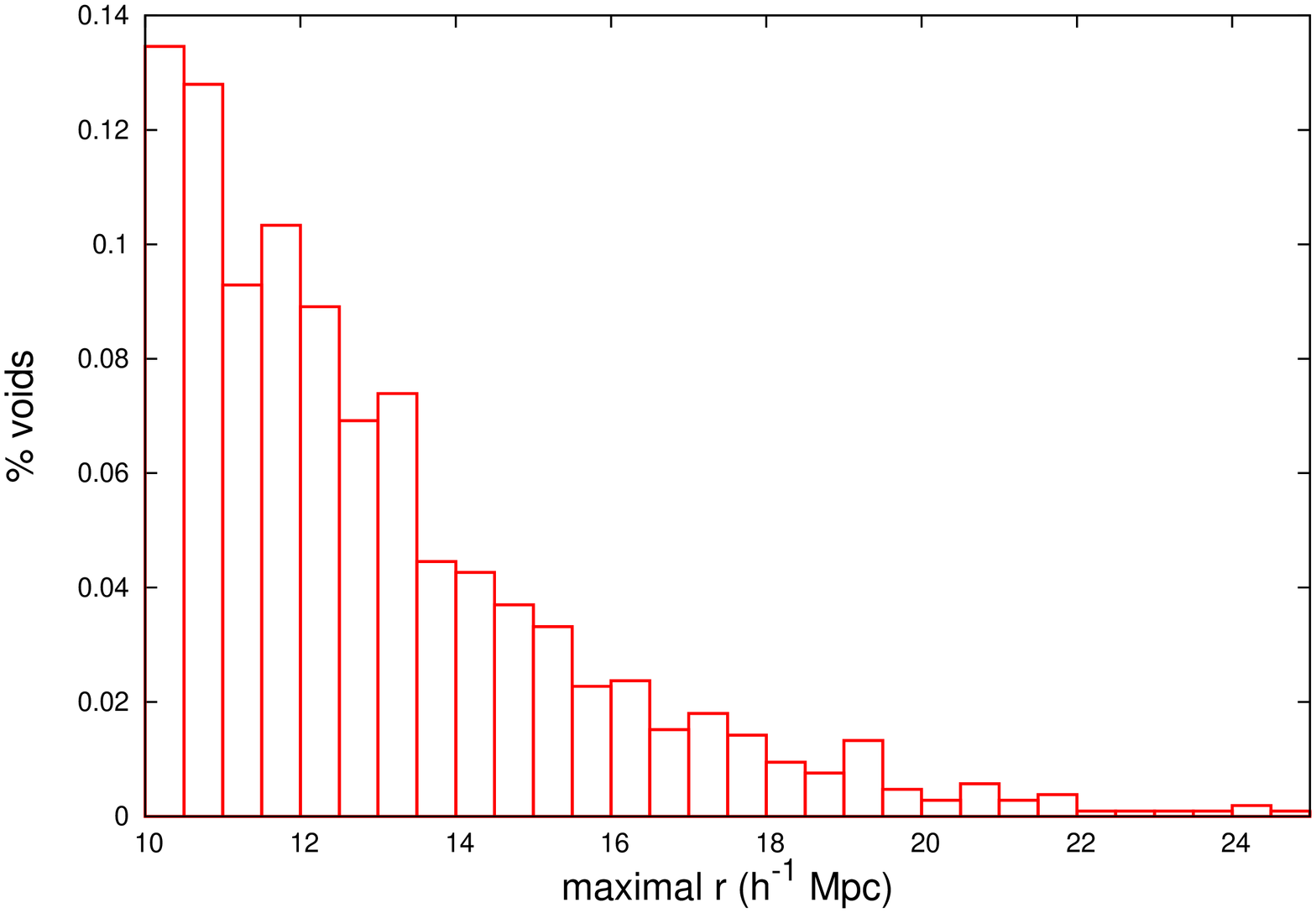}\hfill\includegraphics[scale=0.25,angle=270]{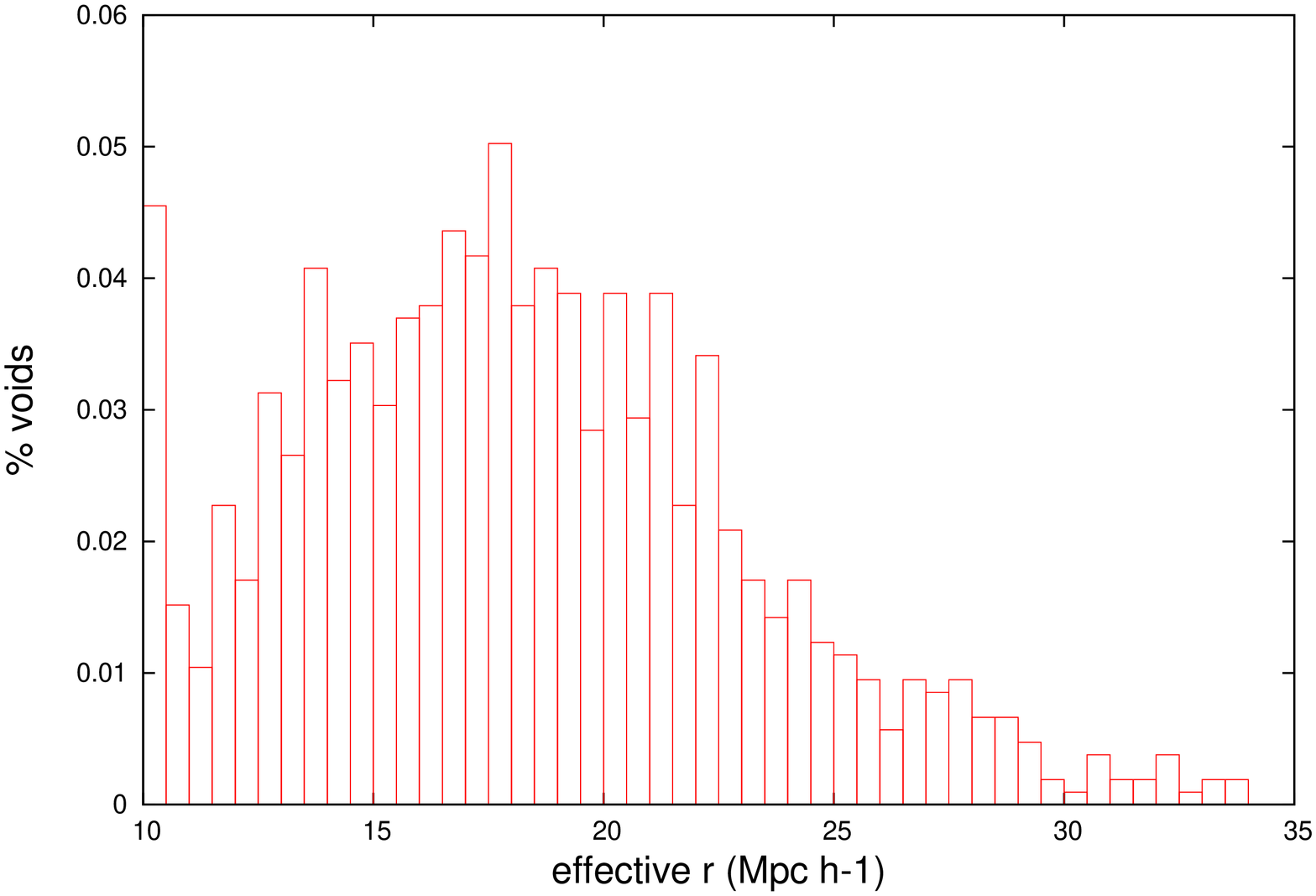}
\caption{Distribution of void sizes as measured by the radius of the maximal enclosed sphere (top panel) 
and by effective radius (bottom panel).
There is a cutoff of 10 $h^{-1}$ Mpc for the holes that make up the voids
and voids with $r_{max}$ near this cutoff make up the majority of the void sample by number.  
The shift in the void distribution from the top to bottom panels indicates that the
the void volumes are not well described by their maximal spheres; most voids are elliptical. Thus, the lack of small voids in the right panel is attributed to their ellipticity.}
\label{radhisto}
\end{figure}

Figure \ref{redshiftslice} shows a redshift slab of 10 $h^{-1}$ Mpc in thickness going through the center of the largest maximal sphere detected by VoidFinder.  The intersections of the plane with all maximal spheres of void regions are shown.  It can be seen that even with just the maximal spheres, a large volume of space is underdense and galaxies cluster strongly in large filament-like structures.  Figure \ref{radhisto} (top) shows the radius histogram based on the largest maximal sphere that defines the void region.  It can be seen that the majority of the voids are small in size with a few very large void regions.  Figure \ref{radhisto} (bottom) shows the effective radius of the individual void regions.  It is important to remember that the maximal spheres are limited to $r > 10 h^{-1}$ Mpc and thus only spherical void regions are found around 10 $h^{-1}$ Mpc in effective radius.  Most voids are not spherical and the skew in the effective radius histogram reflects the ellipticity of the voids.

\begin{figure}
\centering
\includegraphics[scale=0.25,angle=270]{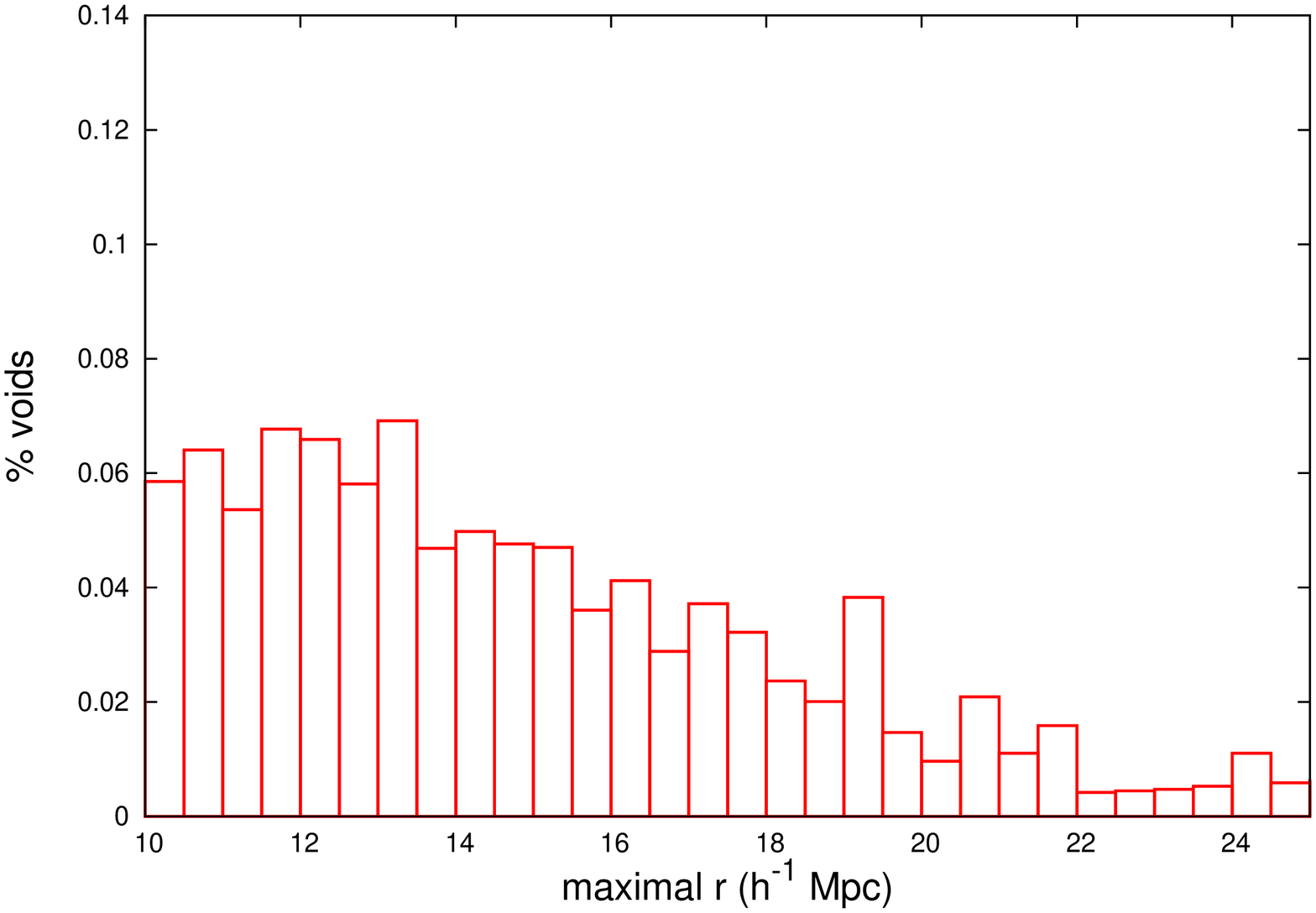}\hfill\includegraphics[scale=0.25,angle=270]{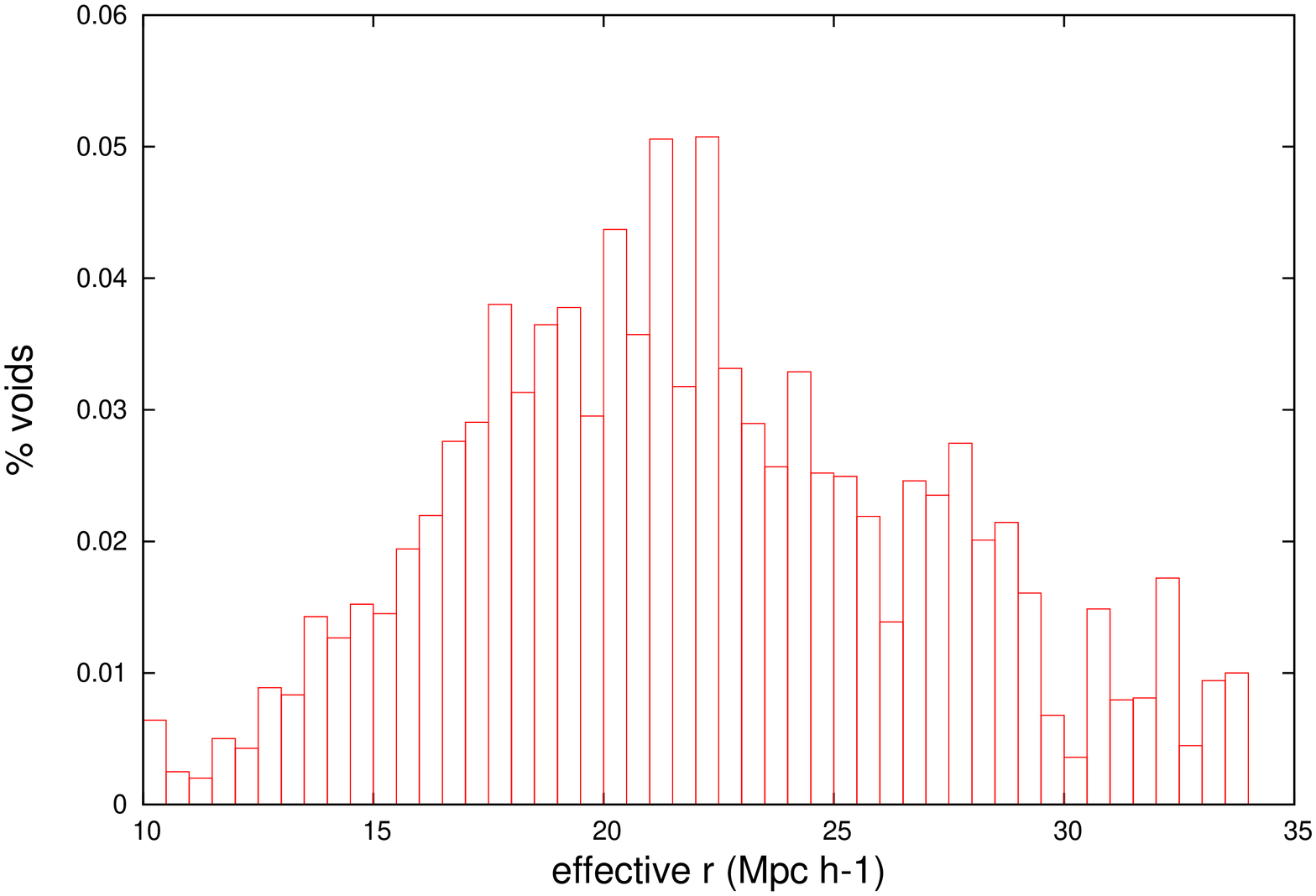}
\caption{Distribution of void sizes as a percentage of the volume occupied by the voids. 
As in Figure 2, the top panel sorts voids by their maximal sphere radii, on the bottom by their effective radii.
Large voids occupy most of the volume with 50\% of the volume occupied by voids with maximal sphere $r > 13.8 h^{-1}$ Mpc, and void size effective $r > 17.8 h^{-1}$ Mpc.  Note the peak of the radius histogram distribution around $22 h^{-1}$ Mpc, the typical size of voids in the Universe. }
\label{radhisto_byvol}
\end{figure}


In figure \ref{radhisto_byvol} we see that the majority of volume occupied by voids are occupied by moderately sized voids.  Even though a large number of voids are smaller in size, the actual volume distribution indicates that there is a preferred size for large scale structure in the Universe.  This is consistent with observations starting with the CfA redshift survey to SDSS.  As indicated in \citet{Shandarin:2006}, void sizes are largely determined by the cosmology.

\subsection{Radial Density Profiles}

\begin{figure}
\centering
\includegraphics[scale=0.5]{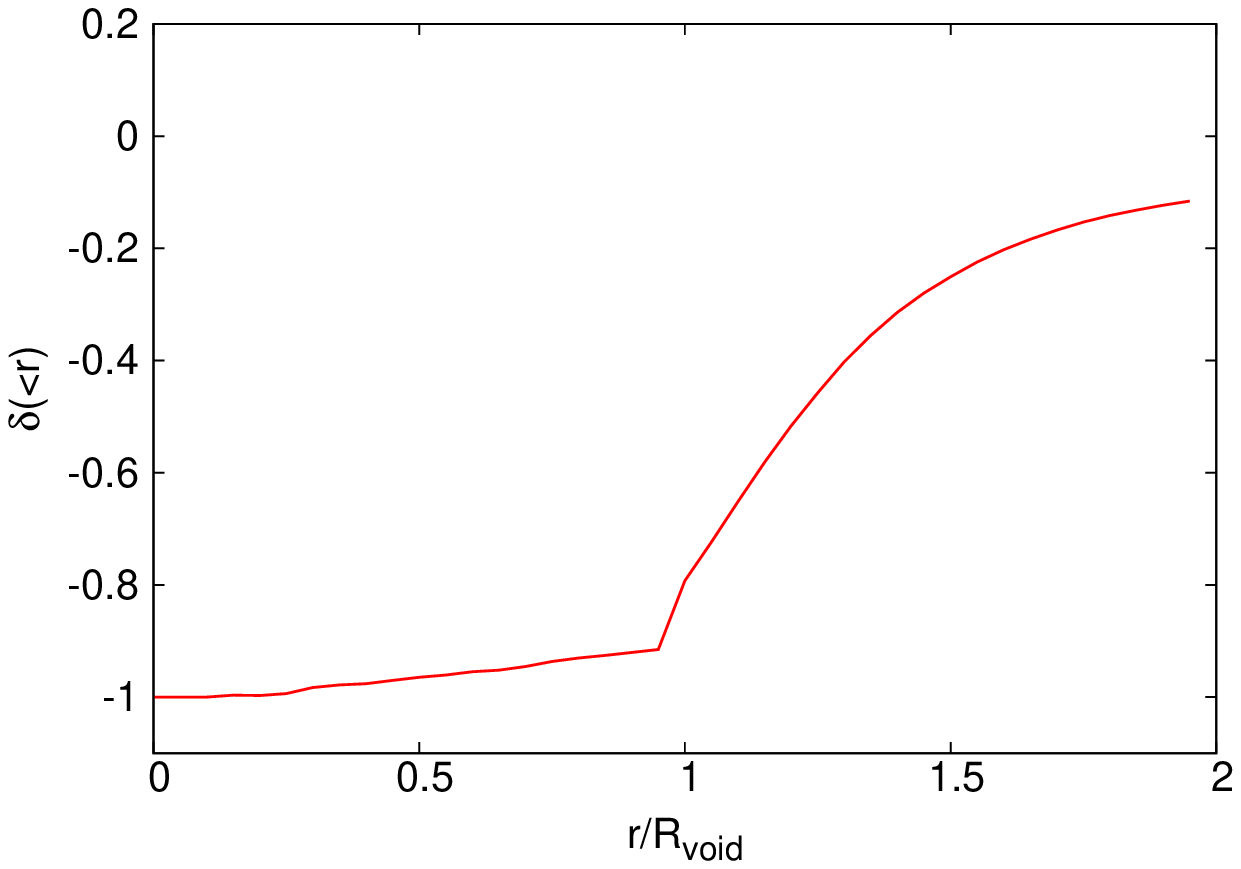}\hfill\includegraphics[scale=0.5]{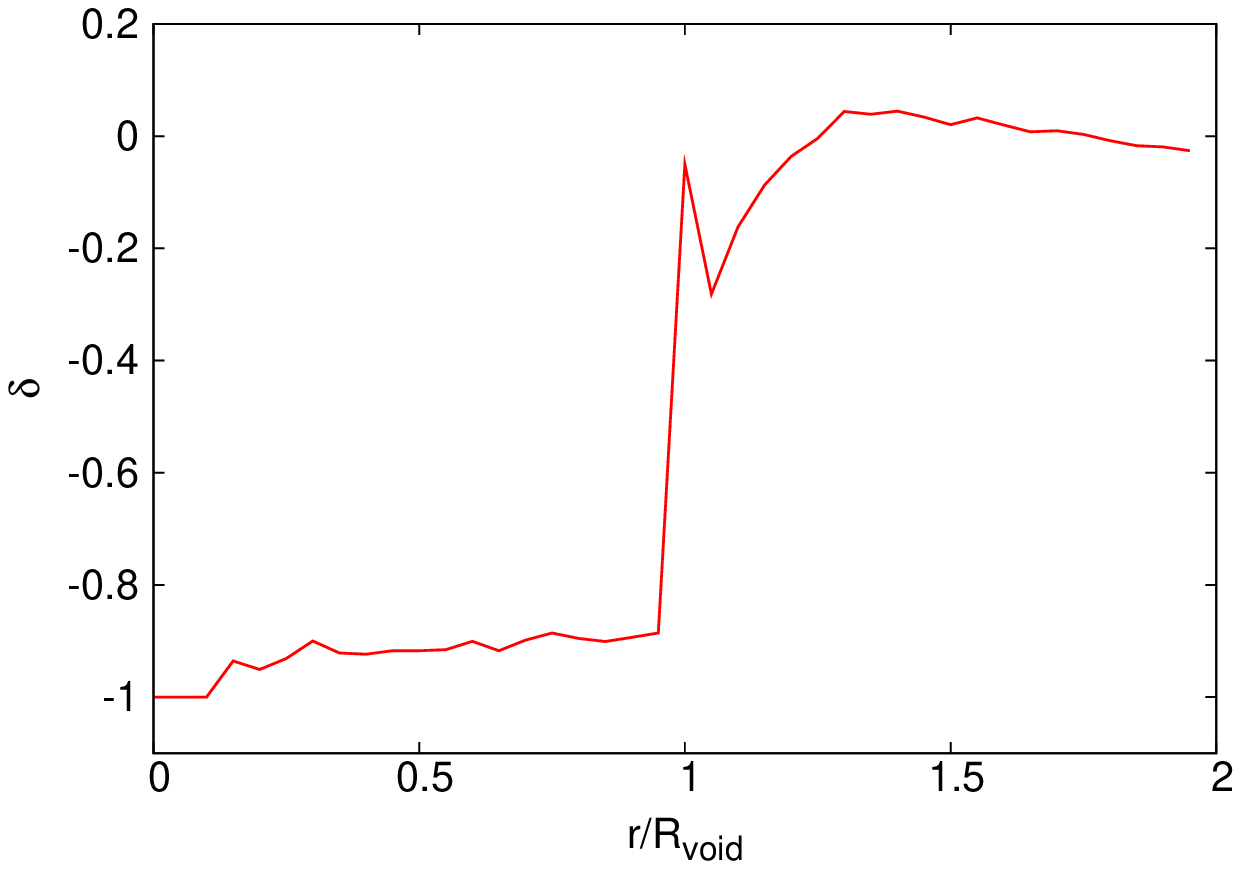}
\caption{The average radial density profile of all 1,054 voids in the void catalog after scaling the profiles by $R_{void}$ and stacking.  The figure on the top is the profile of the enclosed volume, and the figure on the bottom is the profile in spherical shells. In both figures, there is a very sharp spike near the edges of the voids.  The steep rise in the density contrast is because walls of voids are well defined.  The peak at the edge of the void in the spherical shelss may be a feature of the density of the sample.}
\label{radialprofile}
\end{figure}

The radial density profiles of the cosmic voids show that voids are significantly underdense, having less than 10\% of the average density all the way out to the very edge of the voids.  Figure \ref{radialprofile} (top) shows the stacked radial density profile of voids.  The density is calculated from the volume enclosed to the given effective radius of the void.  Figure \ref{radialprofile} (bottom) shows a similar stacked radial density profile of the voids.  However, the density is now calculated for spherical annuli.  It can be seen that the walls of the voids are quite sharp, quickly growing from 10\% of the average density to 100\%, and the voids are very well defined in terms of their density contrast with the outside Universe.  It is clear then that these voids are distinct features of the Universe.  A comparison with linear gravitation theory (Figure \ref{radialprofiletheory}, reproduced from \citet{Sheth:2004}) shows the same ``bucket shaped'' radial density profile (see also Figure 4 of \citet{Fillmore:1984}).

\begin{figure}
\centering
\includegraphics[scale=0.5]{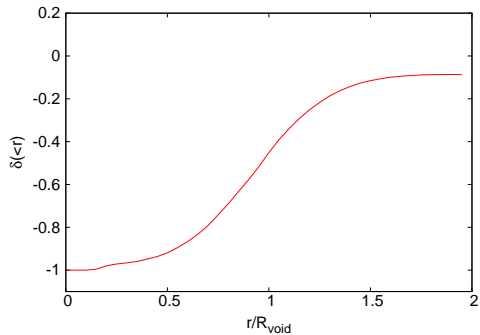}
\caption{The average radial density profile of all 1,054 voids in the void catalog as a function of the effective radius.}
\label{radialprofileer}
\end{figure}

\begin{figure}
\centering
\includegraphics[scale=1]{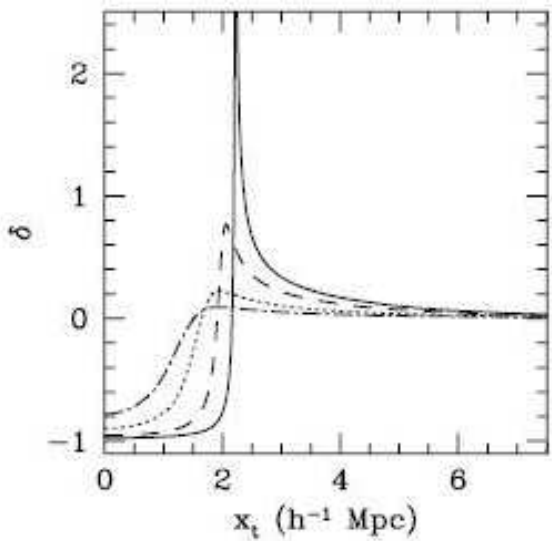}
\caption{Radial density profile (spherical annulus) as predicted by linear gravitation theory \citep{Sheth:2004}.  The different curves correspond to different epochs of evolution, with the tallest peak representing $z = 0$.}
\label{radialprofiletheory}
\end{figure}

\subsection{Void Galaxies}

In our SDSS DR7 galaxy catalog, there are 708,788 galaxies.  In our $M < -20.09$ volume limited galaxy catalog, there are 120,606 galaxies, with 8,046 of them falling inside voids, approximately 7\%.  There are 79,947 (11\%) void galaxies that lie in void regions from the magnitude limited catalog with $z < 0.107$.  Properties of these void galaxies will be discussed in a later paper.

\section{Tests: Volume Limited Cuts}

In this section, we study the effect of changing the absolute magnitude cut on the voids found by VoidFinder.  For absolute magnitudes brighter than $M_r = -20.09$, we use the same redshift cut while eliminating galaxies that fall under the absolute magnitude cut of $-20.2, -20.3...-20.6$.  For absolute magnitudes dimmer than $M_r = -20$, we use a redshift cut of $z = 0.087$, which corresponds to a limiting absolute magnitude of $-19.5$, and apply VoidFinder to samples with magnitude limits of $-19.5, -19.6...-20.1$.  It can be seen that as we slightly shift the absolute magnitude limits the void distribution remains similar, although there are trends that voids generally grow in size with brighter absolute magnitude cuts and voids get smaller in size with dimmer absolute magnitude cuts, as expected for changes in the sampling density of galaxies. 
We find qualitatively different behavior as we examine extremely different samples (L*$\pm$0.5 magnitude), where
we start to observe the effects of merging and splitting of voids. 

\begin{figure}
\centering
\includegraphics[scale=0.25,angle=270]{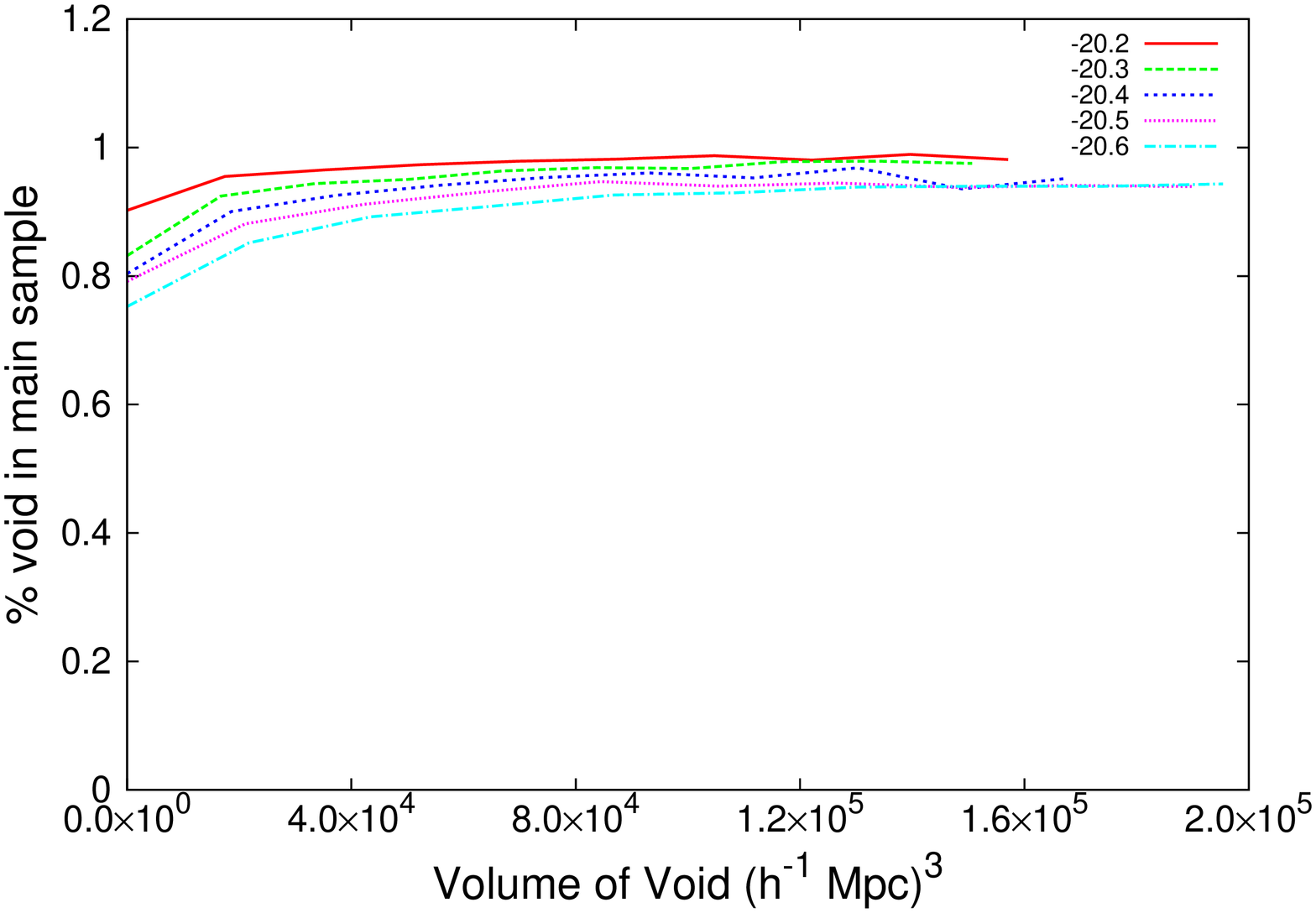}\hfill\includegraphics[scale=0.25,angle=270]{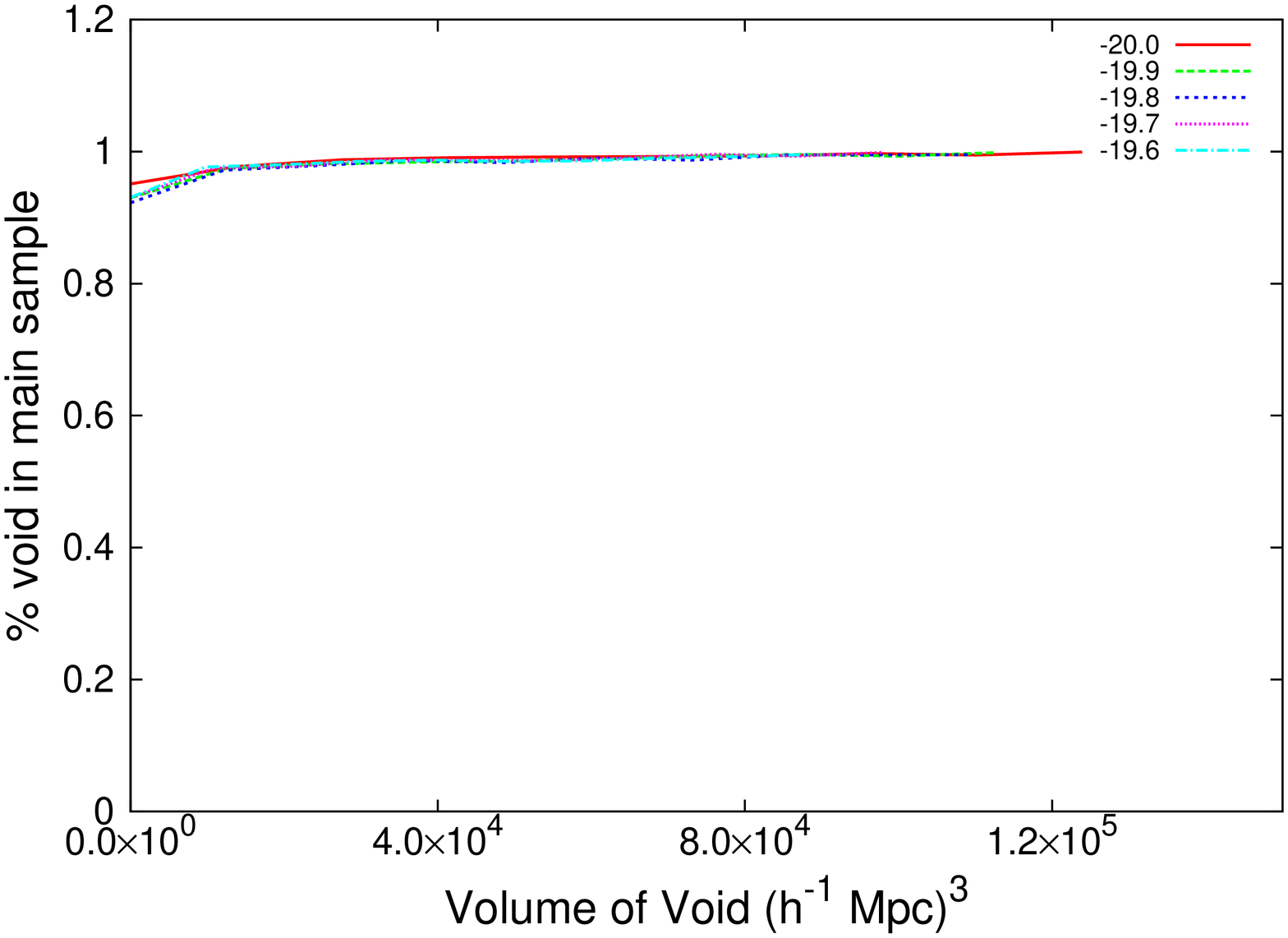}
\caption{Overlap fraction for galaxy samples with redshift cut $z = 0.107$ (top), and $z = 0.087$ (bottom) with magnitude given in the figure compared to the void catalog sample ($M_{lim} = -20.09$).  The y-axis shows the fraction of the void volume that is also considered void in the main sample as a function of the void volume.  It can be seen that the large significant voids are consistently identified regardless of the volume limited cut.}
\label{olap}
\end{figure}
\begin{figure}
\centering
\includegraphics[scale=0.5]{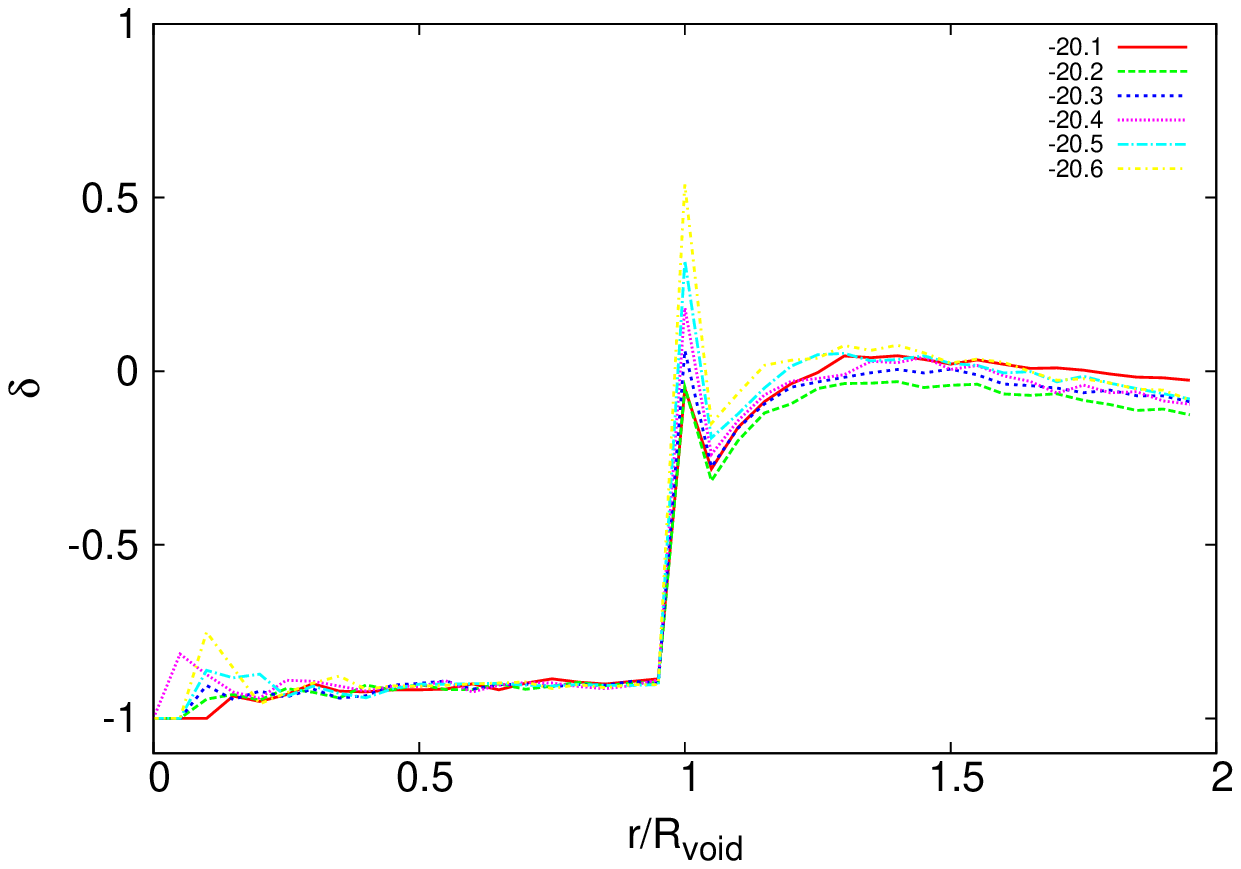}\hfill\includegraphics[scale=0.5]{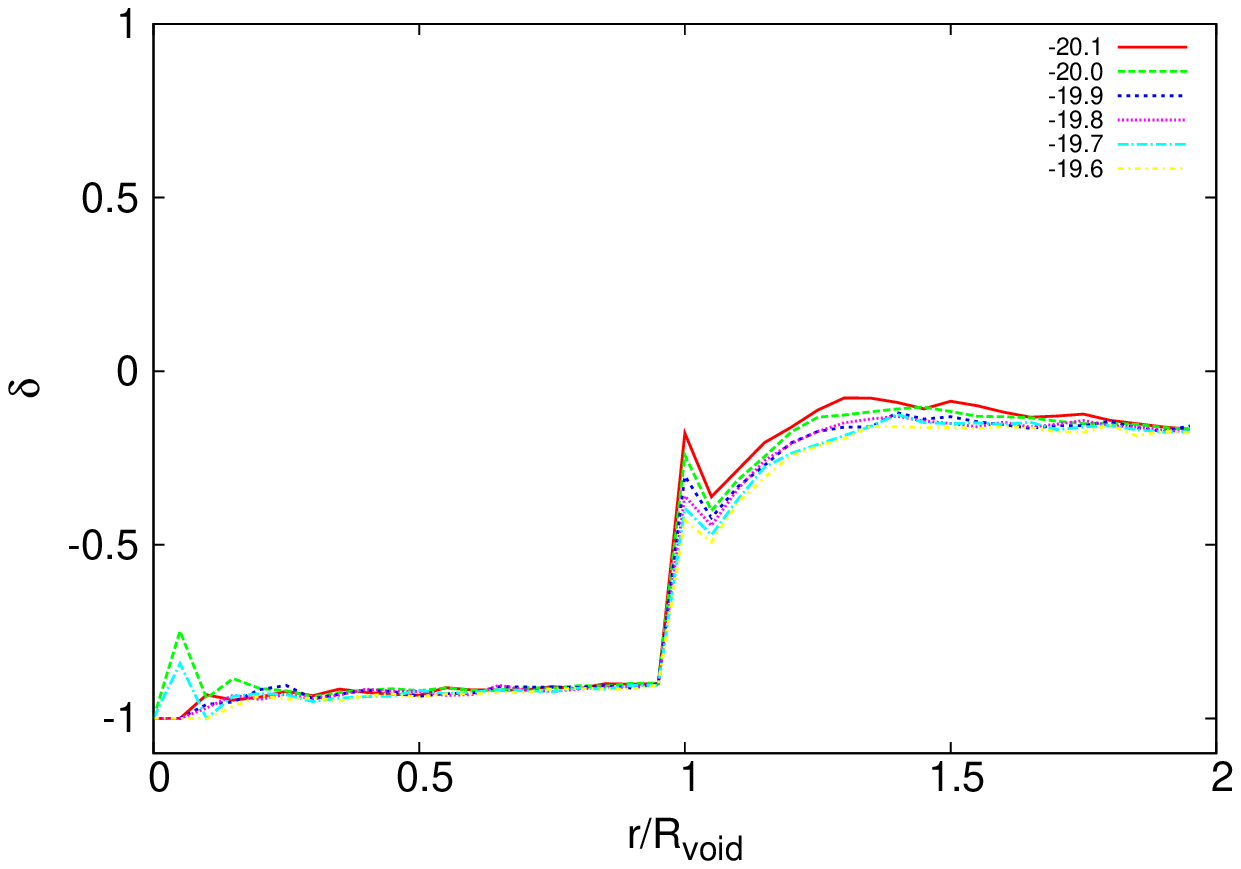}
\caption{Radial density profile (enclosed volume) for galaxy samples with $-20.6 < M_r < -20.1$ (top), and $-20.1 < M_r < -19.6$ (bottom).  The only difference between the profiles is the height of the peak at the edge of the voids.  This is due to the different number density of galaxies in the sample used to determine voids.  The bucket shaped behavior at the walls of the voids is consistent with \citet{Sheth:2004} in Figure \ref{radialprofiletheory}.}
\label{radialprofile_complete}
\end{figure}

Figure \ref{olap} shows that the void regions found by VoidFinder are consistent for almost all large voids.  The only significant discrepancy arises from smaller voids that are introduced in sparser samples of the data.  Figure \ref{radialprofile_complete} shows that the radial density profiles still show the "bucket shaped" feature.

Thus, the voids we find are not very sensitive to the absolute magnitude cut nor to the volume of our sample.  SDSS DR7 provides a sufficiently contiguous three dimensional volume for void finding purposes.  These voids found by VoidFinder should be considered significant large scale underdensities.

\section{Tests: Mock Data}
We test the void finding algorithm on a set of mock galaxy catalogs to analyze the effects of the boundary conditions as imposed by SDSS, the effectiveness of studying large scale 3D structure in the finite volume of SDSS, as well as to test $\Lambda$-CDM predictions of the properties of voids.  The mock catalog used is a SPH halo model \citep{Skibba:2009} enclosed in a cube with sides 480 $h^{-1}$ Mpc.  The luminosity function and luminosity weighted correlation functions of the mock catalogue are fit to SDSS as described by \citet{Skibba:2006}, using halo occupation constraints from \citet{Zheng:2007}.


We test VoidFinder on the mock sample using two different methods.  First, the SDSS mask is applied to the mock sample so that the geometries of the samples are the same; the results of this should mimic that of SDSS DR7.  The SDSS geometry mock catalog contains 98,186 galaxies covering a volume of $2.2\times10^7$ Mpc$^3$ in the volume limited sample.  Second, a cube is selected with volume similar but greater than the SDSS geometry sample.  The cube geometry mock catalog contains 119,076 galaxies covering a volume of $2.7\times10^7 Mpc^3$.


\subsection{Mock Results}
\begin{figure}
\centering
\includegraphics[scale=0.5]{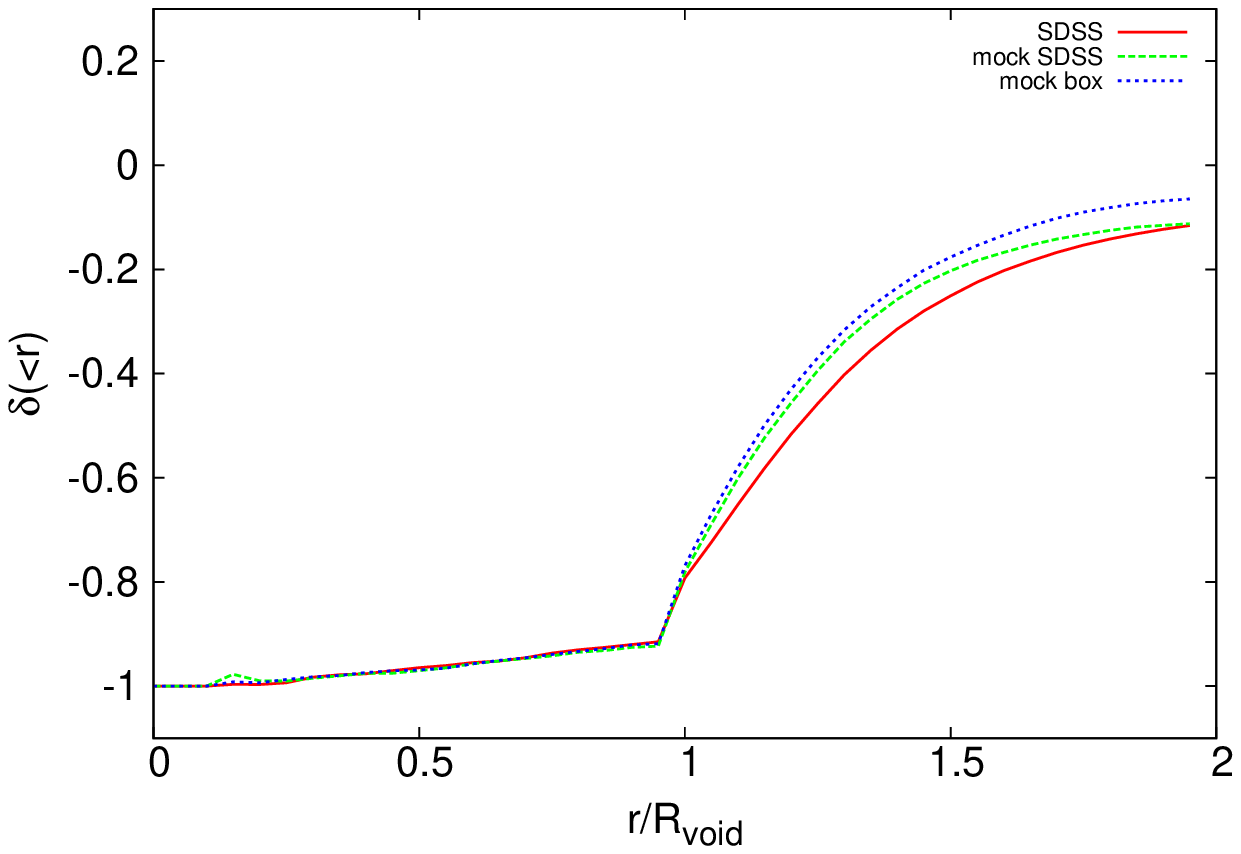}\hfill\includegraphics[scale=0.5]{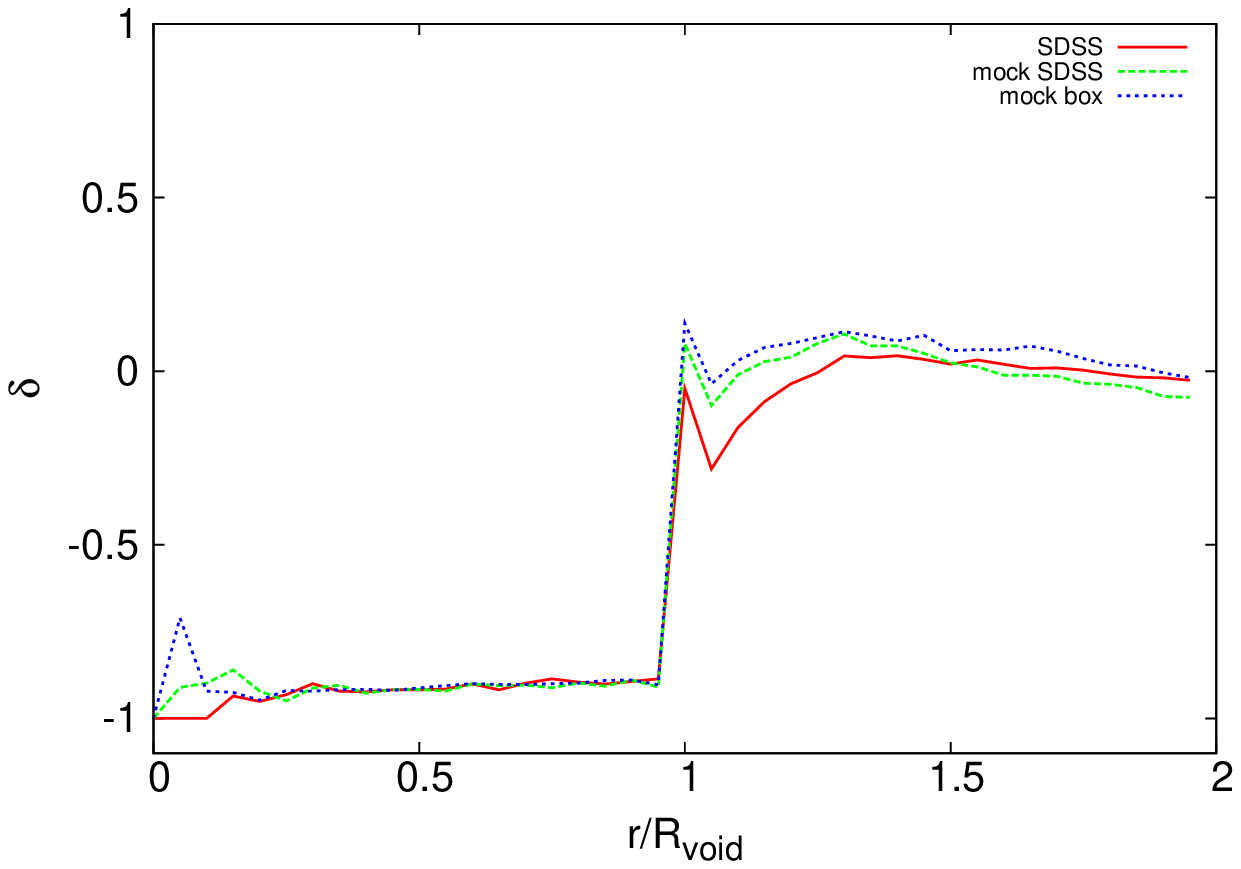}
\caption{Comparison of radial density profiles of voids in SDSS DR7 and simulations (top panel shows enclosed density, bottom panel shows density in spherical shells).  The density profiles within the voids are nearly identical in all cases. The simulations show a slight tendency toward larger density just outside the void boundary.}
\label{radialprofile_mockcompare}
\end{figure}

\begin{figure}
\centering
\includegraphics[scale=0.5]{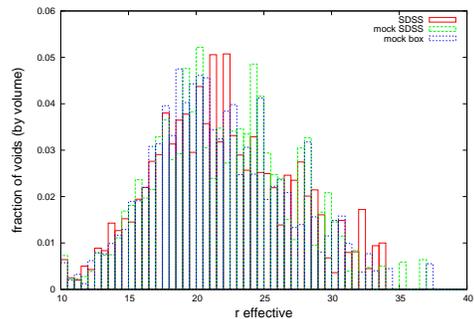}
\caption{Distribution of void filling factor as a function of effective radius for voids found in SDSS DR7 and simulations. 
The distribution of void sizes found in the mock catalogs are nearly identical to those found in SDSS.  The same size voids fill most of the volume.}
\label{radiushisto_mockcompare}
\end{figure}

There are 1,006 voids and 6,228 void galaxies in the SDSS geometry mock catalog.  There are 1,246 voids and 7,881 void galaxies in the cube geometry mock catalog.  The void volume fraction in the SDSS volume cut is 66.5\%, and 69.3\% in the cube volume cut.  We observe that the geometry of SDSS plays a role in determining the overall void volume fraction, and if a SDSS geometry is considered in a mock sample, the volume fraction (66.5\%) is approximately the same as the observed SDSS void volume fraction (62\%).  The effective radius histogram of voids found in the mock samples in Figure \ref{radiushisto_mockcompare} shows no significant changes in the sizes of voids found in the mock samples.  The radial density profile in Figure \ref{radialprofile_mockcompare} shows that the interiors of the voids are similarly empty as well.  The void size and density profile results of the mock samples agree with observational data.  However, there does seem to be a difference in the number of void galaxies found by VoidFinder which will be discussed in a separate paper.


\begin{center}
\begin{tabular}{| c | c | c | c |}
\hline
 & SDSS & SDSS mock & mock cube \\ \hline
\# voids & 1,054 & 1,006 & 1,246 \\ \hline
voids/volume & 0.000048 & 0.000046 & 0.000046 \\ \hline
\# void gals & 8,046 & 6,228 & 7,881 \\ \hline
\# void gals/volume & 0.00037 & 0.00028 & 0.00029 \\ 
\hline
\end{tabular}
\end{center}

\section{Void catalog release}

We have made this void catalog publicly available for future studies of voids.  Included in the catalog are three separate interpretations of void regions.  The first catalog consists of the maximal spheres of each unique void region.  This is the largest hole in each void region in the shape of a sphere.  This catalog is particularly useful for studying vast spherical underdense regions of the Universe.  These spheres often depict the most underdense regions and galaxies near the centers of these voids are living in the most underdense large scale environments.  The second catalog consists of all the possibly overlapping holes identified by VoidFinder.  The merging of the holes forms each unique void region.  This catalog is useful for identifying the entire void distribution of the Universe.  All of the volume enclosed by these holes lies in void regions and all galaxies contained are considered void galaxies.  The third catalog consists of the location and effective size of each unique void region.  This catalog is useful for identifying overall void statistics in the Universe.  Study of large scale structure as well as void volume distributions can be calculated from this catalog.  Along with the three catalogs is the catalog of void galaxies.  We have identified all galaxies with spectra that lie within the void regions identified by VoidFinder.  These catalogs can be downloaded for use\footnote[1]{www.physics.drexel.edu/$\sim$pan/}.

We have now identified the largest and most comprehensive void catalog from the largest spectroscopic data set available.  Previous studies of voids from earlier data releases of SDSS, and other surveys including  2dF Galaxy Redshift Survey all lack the combination of completeness, depth, and contiguous sky provided in SDSS DR7.  There is no longer an issue with survey boundaries restricting the volume of study for finding large voids.  As there are currently no plans for a large spectroscopic survey of L* galaxies, this will be the most comprehensive data set for years to come.


\section{Summary}

We studied the distribution of cosmic voids and void galaxies using Sloan Digital Sky Survey data release 7 using an absolute magnitude cut of $M_r < -20.09$.  Using the VoidFinder algorithm as described by \citet{Hoyle:2002}, we identify 1054 statistically significant voids in the northern galactic hemisphere greater than 10 $h^{-1}$ Mpc in radius, covering 62\% of the volume.  There are 8,046 galaxies brighter than $M_r = -20.09$ that lie within the voids, accounting for approximately 6\% of the galaxies, and 79,947 void galaxies (11.3\%) with $m_r < 17.6$.  The largest void is just over 30 $h^{-1}$ Mpc in effective radii.  The median effective radius is 17 $h^{-1}$ Mpc.  Voids of size $r_{eff} \sim 20h^{-1}$ Mpc dominate the void volume.  The voids are found to be significantly underdense, with $\delta < -0.85$ near the edges of the voids.  We tested the sensitivity of the void finding algorithm to changes in the absolute magnitude cut within the range $-19.6 > M_r > -20.6$.  The resulting void regions are largely similar with slight differences only near the edges of the void regions.  The radial density profiles of the voids are found to be similar to predictions of dynamically distinct underdensities in gravitational theory.  We compared the results of VoidFinder on SDSS DR7 to mock catalogs generated from a SPH halo model simulation as well as other $\Lambda$-CDM simulations and found similar results, ruling out inconsistencies resulting from selection bias and survey geometry.


\bibliographystyle{mn2e}
\bibliography{VoidCatalog_v3}

\end{document}